\documentclass[12pt]{article}
\usepackage{epsfig,amsfonts,amssymb}
\usepackage{hyperref}
\usepackage{comment}
\usepackage{cite}
\input epsf.sty
\usepackage{cite}

\def\ZZZ{{\hbox{ Z\kern-1.6mm Z}}}
\def\RRR{{\hbox{ R\kern-2.4mm R}}}
\def\CCC{{\hbox{ C\kern-2.0mm C}}}
\def\zzz{{\hbox{z\kern-1mm z}}}

\newcommand{\qeq}{{\hbox{=\kern-2.3mm ? \kern.5mm }}}
\renewcommand{\qeq}{=}

\newcommand{\eps}{\epsilon}

\newcommand{\ve}{\varepsilon}

\newcommand{\OO}{{\cal O}}

\newcommand{\be}{\begin{equation}}
\newcommand{\ee}{\end{equation}}
\newcommand{\ben}{\begin{eqnarray}\displaystyle}
\newcommand{\een}{\end{eqnarray}}

\newcommand{\refb}[1]{(\ref{#1})}
\newcommand{\p}{\partial}

\def\one{{\hbox{ 1\kern-.8mm l}}}
\def\zero{{\hbox{ 0\kern-1.5mm 0}}}

\newcommand{\bea}[1]{\begin{eqnarray}\label{#1} }
\newcommand{\eea}{\end{eqnarray}}

\newcommand{\eqref}{\refb}

\usepackage{bm}
\usepackage[table]{xcolor}

\newcommand{\non}{\nonumber}

\begin{document}

\baselineskip 24pt

\begin{center}

{\Large \bf Observational Signature of the Logarithmic Terms in the Soft Graviton Theorem}


\end{center}

\vskip .6cm
\medskip

\vspace*{4.0ex}

\baselineskip=18pt

\centerline{\large \rm Alok Laddha$^{a}$ and Ashoke Sen$^{b}$}

\vspace*{4.0ex}

\centerline{\large \it ~$^a$Chennai Mathematical Institute, Siruseri, Chennai, India}

\centerline{\large \it ~$^b$Harish-Chandra Research Institute, HBNI}
\centerline{\large \it  Chhatnag Road, Jhusi,
Allahabad 211019, India}


\vspace*{1.0ex}
\centerline{\small E-mail:  aladdha@cmi.ac.in, sen@hri.res.in}

\vspace*{5.0ex}

\centerline{\bf Abstract} \bigskip

We show that the recently discovered logarithmic terms in the soft graviton theorem
induce a late time component in the gravitational wave-form that falls off as inverse power
of time, producing 
a tail term to the linear memory effect. 



\vfill \eject

\baselineskip=18pt

One of the reasons for recent interest in soft theorems is its connection to memory 
effect\cite{1411.5745,1502.06120,1502.07644,
1712.01204}
-- the fact that a passing gravitational wave causes a permanent change in the distance between 
two detectors placed in its path\cite{mem1,mem2,mem3,mem4}. 
The connection between soft theorems and memory effect 
usually proceeds via asymptotic symmetries\cite{1312.2229,1401.7026,1408.2228} 
and has led to the prediction of new kind of memory
effect which is associated to the so-called super-rotation symmetries.\cite{1502.06120}.
In contrast,
in \cite{1801.07719} we established a direct connection between soft factors that arise in soft theorems
and low frequency classical radiation in a classical process by taking the classical limit of
the quantum scattering process. 
This has the advantage of being valid in all space-time dimensions, irrespective of whether or not
soft theorems can be related to asymptotic symmetries. However while applying
this formula to four dimensions we encounter a new phenomenon: 
due to long range force on the particles
involved in the scattering, the soft factor at the subleading order gets a contribution proportional to
the logarithm of the energy of the soft radiation\cite{1804.09193}. 
Our goal in this paper will be to describe the observational signature of this logarithmic term in the soft
graviton theorem.
We shall use $\hbar=c=8\pi G=1$ units, although since we are analyzing classical radiation, 
$\hbar$ never enters any formula.

The set up we shall investigate is a process in
which a system of mass $M$, describing the initial system, 
makes a transition into a system of mass $M_0<M$ and some matter / radiation that escapes the
system. We shall work in the rest frame of the original system and
assume for simplicity that the total energy carried by the escaping matter / radiation is small compared 
to the mass of the original system so that $M_0\simeq M$ and 
the recoil velocity of the final
system is small.\footnote{In the analysis of \cite{1804.09193} we had to make this 
assumption since there we
ignored the contribution to the soft factor due to the outgoing 
radiation and therefore had to assume that the
energy carried away by radiation is small compared to that carried by matter. Here that assumption is
not necessary since we shall explicitly take into account the contribution to the soft factor due to the
outgoing radiation. Nevertheless the approximation $M_0\simeq M$ is needed to 
ensure that we can regard the final configuration {\it at late time}
as a set of light particles moving under
the influence of the gravitational field of the stationary heavy object of mass $M_0$, ignoring the 
gravitational force exerted by the outgoing particles on each other. 
More general formula where this 
assumption is not necessary can be derived using the general expression for 
the mutual gravitational force between particles in relative motion\cite{1808.03288}.}
An example of this would be neutron star merger
where large amount of matter is ejected from the parent system but the total amount of energy lost is
still small compared to the mass of the system that remains behind. 
Another example would be
binary black hole
merger, where the energy is carried away by gravitational radiation, but 
we shall see that the effects we shall describe vanish in the case where only massless
particles carry away the energy.
Our focus will be on the low frequency radiative component of the metric field
$h_{\mu\nu}\equiv (g_{\mu\nu}-\eta_{\mu\nu})/2$.

In \cite{1804.09193} a formula for soft radiation was found in a situation where a 
light particle of mass $m$ 
scatters from a heavy particle of mass $M_0$. Here light particle refers to a particle carrying
energy $<< M_0$. However since soft theorem expresses the 
result as independent sum over initial and final states, the result can be easily generalized 
to the case where there are no light particles in the initial state and multiple light particles in
the final state. 
We shall now state this result.

Let $t_0\simeq |\vec x| + M_0 \ln |\vec x| / 4\pi$ 
be the time at which the peak of the gravitational radiation reaches the observer at $\vec x$.
For
the radiative part of the trace reversed metric 
\be \label{edefthij}
\tilde e_{\mu\nu}(\omega, \vec x) \equiv \int dt\, e^{i\omega (t-t_0)} \, e_{\mu\nu}(t,\vec x), 
\quad e_{\mu\nu}(t,\vec x)\equiv \left\{ h_{\mu\nu}(t,\vec x)
-{1\over 2} h_\rho^{~\rho}(t, \vec x) \, \eta_{\mu\nu} \right\}\, ,
\ee
the result of
\cite{1801.07719,1804.09193}, when applied to the situation where we have light particles 
only in the final state, can be
stated as follows.
Up to an overall constant phase that can be absorbed into a shift of the time coordinate, 
we have\footnote{One word of caution may be added here. The general analysis of \cite{1801.07719}, 
on which
this result is based, can determine $\tilde e_{ij}$ up to an overall phase that could depend on the energy
$\omega$ but not on the polarization of the soft graviton. The phase given in \refb{eformmin} can be
derived by comparing the general prediction of soft theorem
with an explicit computation of the radiation during the scattering
of a light particle from.a heavy particle at large impact parameter\cite{1804.09193}. 
We are assuming that the same phase
can be used in this case as well. Since soft radiation comes from the domain where the light particles
have moved sufficiently far away from the heavy center, we expect this relation to be valid. This is also
in keeping with the results of effective field theory approach\cite{0912.4254,1203.2962,1211.6095} where a similar
phase has been found.}
\be \label{eformmin}
\tilde e_{ij}(\omega, \vec x) = {i\over 4\pi \, |\vec x|} \, e^{-iM_0 \,\omega\, \ln\omega^{-1}/(4\pi)} 
\Bigg[\sum_a m_a \, \beta_{ai} 
\, \beta_{aj} {1\over 1-\hat n.\vec \beta_a} {1\over \sqrt{1-\vec \beta_a^2}} \left\{\omega^{-1} 
+ i\, C_a \, \ln\omega^{-1} +\hbox{finite}\right\}
\Bigg]\, ,
\ee
where `finite' refers to terms that have finite $\omega\to 0$ limit,
the sum over $a$ runs over all the light final states, $\vec \beta_a$ is the velocity of the $a$-th
light particle, $m_a$ is the mass of the $a$-th light particle, and
\be \label{e3}
C_a \equiv - M_0 {1-3\vec\beta_a^2\over 8\pi|\vec\beta_a|^3}, \quad \hat n \equiv {\vec x\over |\vec x|},
\quad k\equiv -\omega (1, \hat n)\, .
\ee
The $\tilde e_{0\mu}$ components are undetermined at this stage but can in principle be determined by
using the constraint $k^\mu \tilde e_{\mu\nu}=0$ that follows from linearized Einstein's equation.
The result for $\tilde e_{\mu\nu}$ is ambiguous up to the linearized 
gauge transformation $\delta \tilde e_{\mu\nu}
=\zeta_\mu k_\nu + \zeta_\nu k_\mu - \zeta.k \, \eta_{\mu\nu}$ for any four vector $\zeta$. 

We now define the transverse traceless component $\tilde e^{TT}_{ij}$ as
\be \label{ett1}
\tilde e^{TT}_{ij} = \tilde e_{ij} + \xi_i \, k_j + \xi_j \, k_i - \xi \, \delta_{ij}\, ,
\ee 
where the three vector $\xi_i$ and the scalar $\xi$ are to be chosen such that
\be\label{ett2}
k^i \, \tilde e^{TT}_{ij} =0, \quad \delta^{ij} \tilde e^{TT}_{ij}  = 0\, .
\ee
It is easy to see that $\tilde e^{TT}_{ij}$ is invariant under a gauge transformation. 
Using \refb{eformmin}, \refb{ett1} and \refb{ett2} we now get, after expanding the exponential
factor in \refb{eformmin} to first subleading order,
\ben \label{eform}
\tilde e^{TT}_{ij}(\omega, \vec x) &=& {i\over 4\pi \, |\vec x|} \,   \sum_a  \left\{\omega^{-1} 
+ i\, \left( C_a -{M_0\over 4\pi}\right)\, \ln\omega^{-1} +\hbox{finite}\right\}  m_a \, 
{1\over 1-\hat n.\vec \beta_a} {1\over \sqrt{1-\vec \beta_a^2}}  \left( \beta_{ai}\beta_{aj}\right)^{TT}\, ,
\non\\
\left( \beta_{ai}\beta_{aj}\right)^{TT} &\equiv &
 \left\{\beta_{ai} 
\, \beta_{aj} - {\hat n.\vec \beta_a} (\hat n_i\beta_{aj}+\hat n_j\beta_{ai}) 
+{1\over 2} \left({\vec\beta_a^2} + {(\hat n.\vec\beta_a)^2  } \right)
\hat n_i \hat n_j  + {1\over 2} \left({(\hat n.\vec\beta_a)^2}-\vec\beta_a^2\right)
\, \delta_{ij}
\right\} 
\, .\non\\
\een
The coefficient of $\ln\omega^{-1}$ proportional to $C_a$ represents the effect of late time
radiation from outgoing 
particles accelerating in the background gravitational field, while the term proportional to $M_0/4\pi$ 
represents the effect of backscattering of the soft graviton due to the background gravitational 
field\cite{0912.4254,1203.2962,1211.6095}.

Since \refb{eform} gives the  small $\omega$ behavior of $\tilde e^{TT}_{ij}(\omega, \vec x)$, it
encodes the behavior of its inverse Fourier transform 
$e^{TT}_{ij}(t,\vec x)$ at large time. We shall now explicitly find this behavior.
Let us suppose that $e^{TT}_{ij}$ has the following asymptotic behavior:
\be \label{eq7}
e^{TT}_{ij}(t, \vec x) \simeq \cases{ 0 \quad \hbox{for $u\to-\infty$}\cr
A_{ij} + u^{-1} B_{ij} +\OO(u^{-2}) \quad \hbox{for $u\to\infty$}
}, \quad u\equiv t-t_0\, ,
\ee
where we allow possible logarithmic multiplicative factors in the $\OO(u^{-2})$ terms. As mentioned
earlier, $t_0$ denotes the time when the peak of the gravitational wave reaches the observer at $\vec x$,
but the precise choice is not important since a finite shift will
not affect the expansion coefficients $A_{ij}$ and $B_{ij}$. Since $e^{TT}_{ij}(t,\vec x)$ 
does not fall off as $u\to\infty$, the correct way to interpret \refb{edefthij} is to write
$e^{i\omega u} = (i\omega)^{-1} {d\over du} e^{i\omega u}$ in the Fourier integral and carry out
an integration by parts ignoring boundary terms. This gives
\be 
\tilde e^{TT}_{ij}(\omega, \vec x) = -{1\over i\omega} \int \, du \, 
e^{i\omega u} \, \partial_u e^{TT}_{ij}(t, \vec x)\, .
\ee
We now express this as
\be
\tilde e^{TT}_{ij}(\omega, \vec x) = -{1\over i\omega} \int \, du \, 
\, \partial_u e^{TT}_{ij}(t, \vec x)
-{1\over i\omega} \int \, du \, 
\left\{ e^{i\omega u}-1\right\}  \, \partial_u e^{TT}_{ij}(t, \vec x)\, .
\ee
The first term gives $i\omega^{-1} A_{ij}$. The second term can be estimated by dividing the integration
region to $u<< \eps^{-1}$, $\eps^{-1} < u < \eta \, \omega^{-1}$, $\eta \, \omega^{-1}<u<\omega^{-1}$
and $\omega^{-1} < u<\infty$ where 
$\eps$ and $\eta$ are two small but finite positive numbers. Since
$e^{i\omega u} -1$ is bounded in magnitude by $\omega u$ and since for large negative $u$
$e^{TT}_{ij}(u,\vec x)$ falls off fast, the contribution to $\tilde e^{TT}_{ij}$ 
from the $u<\eps^{-1}$ region remains finite
in the $\omega\to 0$ limit. In the region $\eps^{-1} < u < \eta \, \omega^{-1}$ we can approximate
$e^{i\omega u}-1$ as $i\omega u$ and $\p_u e^{TT}_{ij}$ by $-B_{ij}/u^2$. Therefore
the integral gets a contribution of order $B_{ij} \{\ln \omega^{-1}+ \ln (\eta\,\eps)\}$. 
In the region $\eta \, \omega^{-1}<u<\omega^{-1}$ we can use the results $|e^{i\omega u}-1|\le \omega\, u$
and $\p_u e^{TT}_{ij} \simeq -B_{ij}/u^2$ to argue that the integral is bounded in magnitude by
$\ln (1/\eta)$.
Finally in the
region $u>\omega^{-1}$ we can use the results $|e^{i\omega u}-1|\le 2$ 
and $\p_u e^{TT}_{ij} \simeq -B_{ij}/u^2$ to show that the integral is bounded in magnitude by
$2\, B_{ij}$. Therefore by taking $\eps,\eta$ to be small but fixed, we see
that the leading contribution to the integral for
small $\omega$ is given by
\be \label{ecomp1}
\tilde e^{TT}_{ij}(\omega,\vec x) = i\, \omega^{-1} \, A_{ij} + B_{ij}\, \ln\omega^{-1} + \hbox{finite}\, .
\ee
Comparing \refb{eform} and \refb{ecomp1}, and using \refb{e3},
we can determine the coefficients $A_{ij}$ and $B_{ij}$:
\be \label{eaaij}
A_{ij} = {1\over 4\pi \, |\vec x|} \,   \sum_a
m_a \, 
{1\over 1-\hat n.\vec \beta_a} {1\over \sqrt{1-\vec \beta_a^2}}  \left( \beta_{ai}\beta_{aj}\right)^{TT}
= {2\, G\over |\vec x|} \,   \sum_a
m_a \, 
{1\over 1-\hat n.\vec \beta_a} {1\over \sqrt{1-\vec \beta_a^2}}  \left( \beta_{ai}\beta_{aj}\right)^{TT}\, ,
\ee
\ben \label{ebbij}
B_{ij} &=& {M_0\over 32\pi^2 \, |\vec x|} \,   \sum_a
m_a \, 
{1\over 1-\hat n.\vec \beta_a} {1\over \sqrt{1-\vec \beta_a^2}} 
{1-3\vec\beta_a^2 + 2|\vec\beta_a|^3 \over |\vec\beta_a|^3} \left( \beta_{ai}\beta_{aj}\right)^{TT}\nonumber \\
&=& {2\, G^2\, M_0\over |\vec x|} \,   \sum_a
m_a \, 
{1\over 1-\hat n.\vec \beta_a} {1\over \sqrt{1-\vec \beta_a^2}} 
{1-3\vec\beta_a^2 + 2|\vec\beta_a|^3 \over |\vec\beta_a|^3} \left( \beta_{ai}\beta_{aj}\right)^{TT}\, ,
\een
where in the last steps we have rewritten the result in terms of the Newton's constant $G=1/8\pi$.

If we consider the case where all the final states
are massless / ultra-relativistic particles for which $|\vec \beta_a|=1$, 
then we see from \refb{ebbij} that $B_{ij}$ vansihes. Therefore in this case there will be no tail effect
-- showing that non-linear memory effect\cite{christodoulou,thorne,1003.3486,1401.5831,1312.6871} 
has no tail term of order $1/u$.
This shows that in order to realize the tail effect we need to focus on processes where some of the final
state light particles are massive, {\it e.g.} in the merger of two neutron stars.
For small $|\vec\beta_a|$ 
the tail term appears to dominate over the memory term, -- in fact \refb{ebbij} gives the
impression that $B_{ij}$ diverges as $|\vec\beta_a|\to 0$. However we note that there is no real singularity in the 
$|\vec\beta_a|\to 0$
limit since it will take a period of order $G M_0/|\vec\beta_a|^3$ for the kinetic energy $m_a\vec\beta_a^2/2$
of the particles to 
dominate the potential energy $GM_0m_a/(|\vec\beta_a|u)$ so that we can use the asymptotic formula for
the particle trajectory used in deriving \refb{eformmin}. After waiting for time $u\sim
G M_0/|\vec\beta_a|^3$  
after the peak of the gravitational wave has passed, the tail term $B_{ij}/u$ already becomes of the order of  
the memory term $A_{ij}$. It falls below the memory term as $u$ increases further.

The tail effect has been discussed in various contexts {\it e.g.}
in \cite{blanchet,blanchet1,blanchet2,9710038,0812.0069,0902.3660,1004.4209,1702.06839}, 
and is presumably related to the effect discussed here. Eqs.~\refb{eq7}, \refb{ebbij} show that 
soft theorem provides a way to express
this in a compact form in terms of the velocity and mass distribution of the outgoing particles.

\bigskip

{\bf Acknowledgement:}
We would like to thank P. Ajith, K.G.~Arun, Luc Blanchet, Miguel Campiglia, Thibault Damour, Bala Iyer,
Ira Rothstein, Walter Goldberger and Biswajit Sahoo
for useful discussions. 
A.S. would like to thank the Abdus Salam International Centre for Theoretical Physics for
hospitality during the final stages of this work. A.L. would like to thank International Center for Theoretical Sciences for their hospitality during the final stages of the work.  The work of A.S. was
supported in part by the J. C. Bose fellowship of 
the Department of Science and Technology, India.

\end{document}